\documentclass[superscriptaddress, reprint,altaffillsymbol]{revtex4-2}
\usepackage{graphicx} 
\usepackage{gensymb} 
\usepackage{mhchem} 
\usepackage{xcolor} 

\usepackage{setspace}
\usepackage{caption}
\usepackage{subcaption}

\begin{document}
    \title{Narrow optical linewidths in stoichiometric layered rare-earth crystals} 

    \author{Donny R. Pearson Jr.}
    \altaffiliation{These authors contributed equally to this work.}
        \affiliation{Department of Physics, University of Illinois Urbana-Champaign, Urbana, IL 61801}
        \affiliation{Materials Research Laboratory, University of Illinois Urbana-Champaign, Urbana, IL 61801}

    \author{Ashwith Prabhu}
    \altaffiliation{These authors contributed equally to this work.}
        \affiliation{Department of Physics, University of Illinois Urbana-Champaign, Urbana, IL 61801}
        \affiliation{Materials Research Laboratory, University of Illinois Urbana-Champaign, Urbana, IL 61801}
        
    \author{Selvin Tobar}
        \affiliation{Department of Chemical Engineering, University of Illinois Urbana-Champaign, Urbana, IL 61801}
        \affiliation{Materials Research Laboratory, University of Illinois Urbana-Champaign, Urbana, IL 61801}
        
    \author{Jack D'Amelio}
        \affiliation{Department of Chemistry, University of Illinois Urbana-Champaign, Urbana, IL 61801}
        \affiliation{Materials Research Laboratory, University of Illinois Urbana-Champaign, Urbana, IL 61801}
        
    \author{Amy Tram}
        \affiliation{Lake Forest College, Lake Forest, IL 60045}
        \affiliation{Materials Research Laboratory, University of Illinois Urbana-Champaign, Urbana, IL 61801}

    \author{Zachary W. Riedel}
     \altaffiliation{Present Address: Los Alamos National Laboratory, Los Alamos, NM 87545}
     \affiliation{Department of Materials Science and Engineering, University of Illinois Urbana-Champaign, Urbana, IL 61801}
        \affiliation{Materials Research Laboratory, University of Illinois Urbana-Champaign, Urbana, IL 61801}

    \author{Daniel P. Shoemaker}
        \affiliation{Department of Materials Science and Engineering, University of Illinois Urbana-Champaign, Urbana, IL 61801}
        \affiliation{Materials Research Laboratory, University of Illinois Urbana-Champaign, Urbana, IL 61801}
        
    \author{Elizabeth A. Goldschmidt}
    \email{goldschm@illinois.edu}
        \affiliation{Department of Physics, University of Illinois Urbana-Champaign, Urbana, IL 61801}
        \affiliation{Materials Research Laboratory, University of Illinois Urbana-Champaign, Urbana, IL 61801}

\begin{abstract}
    Rare-earth emitters in solids are well-suited for implementing efficient, long-lived quantum memory coupled to integrated photonics for scalable quantum technologies. They are typically introduced as dopants in a solid-state host, but this introduces disorder and limits the available density of emitters. Stoichiometric materials can offer high densities with narrow optical linewidths. The regular spacing of emitters also opens possibilities for quantum information processing and collective effects. Here we show narrow optical linewidths in a layered stoichiometric crystalline material, NaEu(IO$_3$)$_4$. We observed an inhomogeneous linewidth of 2.2(1)~GHz and a homogeneous linewidth of 120(4)~kHz. Using spectral hole-burning techniques, we observe a hyperfine spin lifetime of 1.9(4)~$\rm{s}$. Furthermore, we demonstrate an atomic frequency comb delay of up to 800~ns.
\end{abstract}

\maketitle

    Optical quantum memories, in which a photonic qubit is stored in an atomic ensemble or other device, have wide potential uses for quantum information technologies \cite{simon2010quantum}. This includes the realization of quantum repeaters to enable the distribution of entanglement over long distances connected by lossy channels as well as synchronization of photonic qubits in quantum networks or local quantum optical systems \cite{sangouard2011quantum, boone2015entanglement}. Other proposed uses for quantum sensing and metrology and more complex quantum information schemes have been proposed \cite{bussieres2013prospective, goldner2015rare}. Different schemes requiring quantum memories have different requirements on metrics such as efficiency of storage and retrieval, storage time, acceptance bandwidth, and more. Rare-earth emitters in solids have many fundamental properties that make them promising for developing scalable, long-lived, and efficient quantum memory devices \cite{guo2023rareearth, shinbrough2023broadband, zhong2019emerging}. Their solid-state nature eliminates the motional dephasing that plagues many atomic gas-based quantum memories and invites the potential for photonic integration \cite{miyazono2017coupling, craiciu2019nanophotonic, sinclair2017properties}. Their excellent coherence properties and high emitter densities are good for achieving long storage times and high storage efficiencies \cite{ma2021onehour, hain2022fewphoton, schraft2016stopped}. 
    
    A major outstanding challenge with rare-earth materials is combining all the desirable properties in a single system due to the inability to resolve the longest-lived spin transitions over typical inhomogeneous linewidths and the reduced coherence observed for rare-earths embedded in materials compatible with standard nanofabrication techniques. One promising pathway for long-lived and efficient RE-based quantum memories is the use of stoichiometric or molecular crystals where the lack of intentional doping can lead to a more homogeneous environment and less disorder-induced broadening, while also achieving extremely high emitter densities \cite{ahlefeldt2016ultranarrow, serrano2022ultra}. To be useful for quantum memory and other applications the RE emitters in these materials will ideally have narrow homogeneous linewidths, narrow inhomogeneous linewidths, and long spin lifetimes. 
    
    Here we present a stoichiometric europium crystal, \ce{NaEu(IO3)4}, that exhibits a homogeneous linewidth of 120~kHz, an inhomogeneous linewidth of 2.2~GHz, and a spin lifetime of more than 2~s. Furthermore, it is a two-dimensional layered material, which opens up exciting possibilities for hybrid photonic integration with high-quality nanophotonics \cite{oh2017photoconversion, youngblood2017integration, meng2023functionalizing, cui2022chip}. Several molecular materials have recently been investigated for QIP and are an attractive platform due to the ability to engineer the material's physical properties by designing and tuning the ligands \cite{kumar2021optical,schlittenhardt2024spectral}. The range of homogeneous and inhomogeneous optical linewidths and spin lifetimes in these crystals is similar to the results presented here \cite{kumar2021optical, serrano2022ultra, kuppusamy2023observation, schlittenhardt2024spectral}. 
    There is one stoichiometric Eu-containing crystal that has been shown to have sufficiently small optical inhomogeneous linewidth to resolve the individual hyperfine spin transitions, EuCl$_3\cdot6$H$_2$O~\cite{ahlefeldt2016ultranarrow}; however, its hygroscopic nature drives the ongoing search for a material with all the desired optical and spin properties that are also air and vacuum stable.

\begin{figure*}
    \centering
    \includegraphics[width=\textwidth]{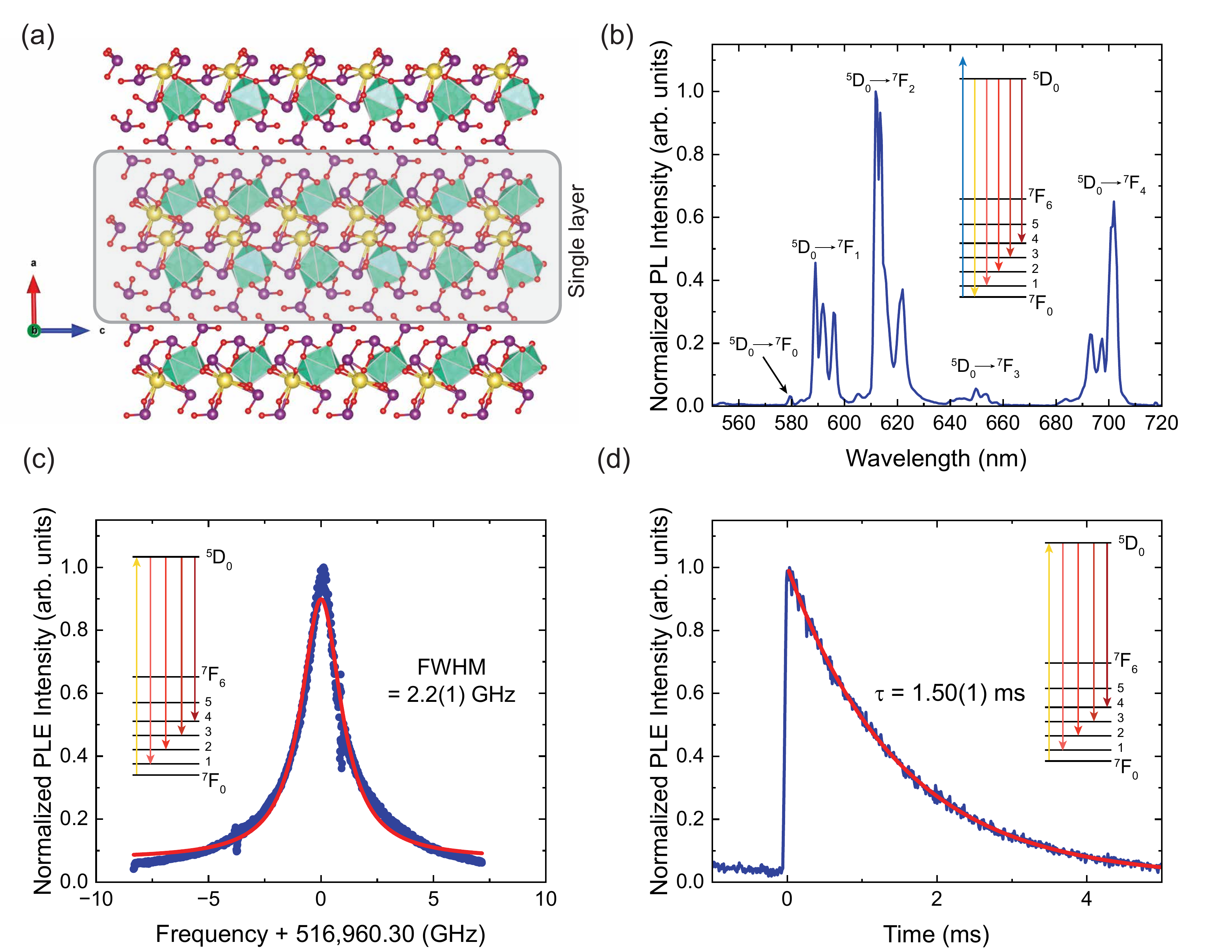}
    \caption{(a) A ball-and-stick model of \ce{NaEu(IO3)4} viewed in the $ac$-crystallographic plane (ICSD entry: 244007 \cite{oh2017photoconversion}). A single monolayer is represented by the gray box. The Eu$^{3+}$ coordination environment with the eight oxygen atoms is given by the teal polyhedra [Eu: teal, Na: yellow, I: violet, O: red]. (b) Photoluminescence spectrum was taken at 1.7~K with 405~nm excitation light. (c) Photoluminescence excitation spectrum taken at 1.7~K. (d) Photoluminescent lifetime at 1.7~K taken using 50~$\mu$s resonant excitation pulses.} 
    \label{fig:PLplots}
\end{figure*}

\ce{NaEu(IO3)4} is an environmentally stable crystalline material that we produced by hydrothermal synthesis to obtain rod-shaped crystals as large as $\approx$ 0.1-0.3~mm in the longest dimension. It crystallizes in the monoclinic crystal system with the noncentrosymmetric space group $Cc$ (see Fig. \ref{fig:PLplots}a). The Eu$^{3+}$ ions are surrounded by eight oxide ligands in a distorted square antiprismatic coordination environment \cite{ok2005new, oh2017photoconversion}. The nearest-neighbor Eu-Eu distance is 5.6~\AA~along the $b$-crystallographic axis, large enough to mitigate the effect of direct Eu-Eu interactions. \ce{NaEu(IO3)4} has a high density of Eu ions at $3.6 \times 10^{21}~\rm{cm}^{-3}$.  No isotopic purification of the host atoms is required as \ce{^127I} and \ce{^23Na} are the only stable and abundant isotopes of iodine and sodium, and \ce{^16O} occurs in an abundance of over 99.7\%. We note that \ce{^127I} and \ce{^23Na} have nuclear spins of $I = + \frac{5}{2}$ and $+ \frac{3}{2}$, respectively, which likely have a negative impact on the spin coherence time, but Eu$^{3+}$ has been shown to have extremely long spin coherence times even in materials with abundant nuclear spins \cite{zhong2015optically, ahlefeldt2013optical, shelby1980frequency}. 

Figure \ref{fig:PLplots}b shows the photoluminescence (PL) spectrum of \ce{NaEu(IO3)4} at 1.7~K under excitation at 405~nm. The spectrum shows clearly the $^5$D$_0$$\rightarrow$$^7$F$_J$ emission lines for $J=0,1,2,3,4$. Additional peaks at $J=5, 6$ can be seen at room temperature (see supplementary information). We focus on the $^5$D$_0$$\rightarrow$$^7$F$_0$ transition at 579.914~nm, which exhibits narrow homogeneous linewidths in Eu-doped crystals. We perform high-resolution photoluminescence excitation (PLE) spectroscopy of this transition in single crystals at 1.7~K. We excite resonantly on the $^5$D$_0$$\rightarrow$$^7$F$_0$ transition and collect the emission via a confocal microscope. We spectrally filter out the excitation light and detect the red-shifted fluorescence on the $^5$D$_0\rightarrow^7$F$_{J>0}$ transitions with a silicon single photon avalanche diode (SPAD). Figure \ref{fig:PLplots}c shows the photon counts as a function of excitation wavelength around the central wavelength 579.914~nm (516960.3~GHz). We observe an inhomogeneously broadened absorption line with a 2.2(1)~GHz linewidth extracted from a Lorentzian fit. This inhomogeneous linewidth is similar to those reported in other low disorder, highly crystalline stoichiometric materials such as [Eu$_2$], Eu(tresnal), Eu(BA)$_{4}$](pip), [Eu(dpphen)(NO$_{3}$)$_{3}$], and EuP$_5$O$_{14}$ \cite{kuppusamy2023observation, kumar2021optical, serrano2022ultra,schlittenhardt2024spectral, shelby1980frequency}, albeit broader than the extremely narrow inhomogeneous linewidths seen in  Eu$^{35}$Cl$_3$$\cdot$6H$_{2}$O \cite{ahlefeldt2016ultranarrow}.

We measure the lifetime of the $^5$D$_0$ state by sending in short (50~$\mu\rm{s}$) resonant pulses and detecting the exponentially decaying red-shifted fluorescence on a SPAD (see Fig. \ref{fig:PLplots}d). We observe a 1.50(1)~ms lifetime,  longer than several previously studied molecular Eu crystals \cite{schlittenhardt2024spectral, kuppusamy2023observation, serrano2022ultra}, though comparable to two metal-organic frameworks we previously reported \cite{riedel2022synthesis}, and shorter than stoichiometric EuCl$_3$$\cdot$6D$_{2}$O (2.6~ms) \cite{ahlefeldt2013optical} and many Eu-doped crystals \cite{ferreira2012dependence}. Similarly, we extract the oscillator strength, which parameterizes the light-matter interaction \cite{thiel2011rare}, for the \ce{^5D_0}$\rightarrow$\ce{^7F_0} transition from the PL and lifetime measurements. We find it to be $7.4(1)\times10^{-9}$, similar to values for other Eu$^{3+}$ materials \cite{thiel2011rare,shelby1980frequency, tyminski1982energy,schlittenhardt2024spectral,ahlefeldt2013optical,serrano2022ultra}.

\begin{figure*}
    \centering
    \includegraphics[width=\textwidth]{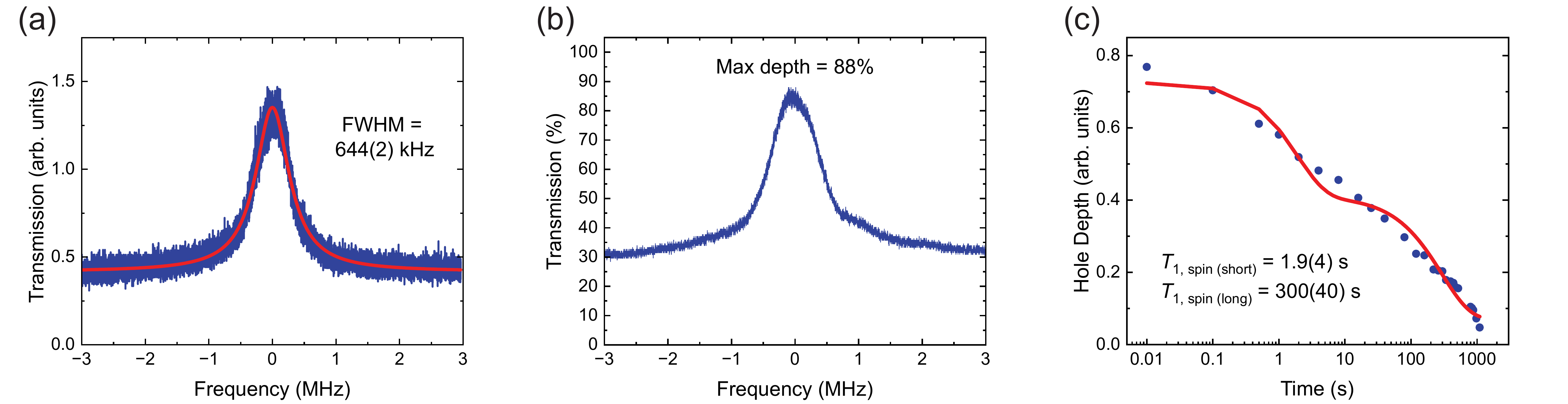}
    \caption{Spectral hole-burning studies in \ce{NaEu(IO_{3})_{4}} at 1.7~K. (a) A narrow spectral hole is observed by transmission measurement. The spectral hole was fit to a Lorentzian function given by the red line. (b) A spectral hole with the maximum achieved transmission of 88\%. (c) The spectral hole depth as a function of the delay time between the burn and probe pulse. The experimentally measured data was fit to a bi-exponential given by the red line.}
    \label{fig:SHB}
\end{figure*}

The ability to spectrally tailor the ensemble is necessary for implementing quantum memory and other protocols in rare-earth ensembles. In particular, burning narrow, deep, and long-lived spectral features is required to optically polarize (or otherwise tailor) the hyperfine spin populations. We show such spectral hole-burning and also use the results to gain further insight into the spin and optical properties of the Eu ensemble in \ce{NaEu(IO3)4}. This was done in a transmission geometry on a densely packed powder because our single crystals were not large enough for a focused beam to propagate through without substantial optical scattering. We crushed several crystals into a powder, which was packed into a small volume formed by a 4~mm diameter and 0.75~mm thick circular aperture sandwiched between two sapphire wafers. We note that this sample exhibited a larger inhomogeneous linewidth of 7.6(5)~GHz (see Supplementary Information), likely due to variations between microcrystals and added strain. We hope soon to be able to grow sufficiently large crystals to enable transmission measurements without overwhelming scatter from surfaces. A bright, single-frequency optical field illuminated the sample for between 100~ms and 800~ms, optically pumping Eu$^{3+}$ emitters into metastable hyperfine states where the transition energy no longer matches the pump frequency, resulting in increased transmission at the pump frequency. Following a 10~ms waiting period to allow all emitters to decay, a weak probe field was scanned across the hole-burning frequency, and the transmitted light was detected with an avalanche photodiode. In Figure \ref{fig:SHB}a we see a peak in the transmission at the hole-burning frequency that is well-fit to a Lorentzian of width 644(2)~kHz. The hole half-width represents an upper bound on the homogeneous linewidth of the Eu emitters with the true linewidth determined more accurately with photon echo measurements described below. We expect that this hole width is broadened by the laser linewidth during hole burning and probing. A maximum hole depth of 88~\% was achieved 2~GHz off-resonance where the optical depth is $\approx 1.5$, compared to an optical depth of $\approx 2$ at the center of the inhomogeneous profile. We used burn pulses of 1~$\rm{s}$ (see Fig. \ref{fig:SHB}b), sacrificing the narrow width achieved with shorter burn times for higher transmission. The ability to fully remove population from a spectral region is important for achieving high-efficiency storage. Finally, we observed biexponential decay of the depth of the spectral hole as a function of time after the burn with a short timescale of 1.9(4)~s and a long timescale of 300(40)~s (see Fig. \ref{fig:SHB}c). This multi-exponential decay is expected for the redistribution of the hyperfine ground states due to the different rates of relaxation mechanisms such as inelastic Raman scattering, direct-phonon processes, and multi-phonon processes \cite{konz2003temperature, macfarlane1987coherent, thiel2011rare}. 

The optical coherence time/homogeneous linewidth determines the limit on how long a photon could be stored on the optical transition, how sharp spectral features can be generated via spectral hole burning, and how much excess broadening would be seen on emission from individual ions. In europium doped oxides, the homogeneous linewidth can be $<200~\rm{Hz}$ and within a factor of two of the lifetime limit \cite{equall1994ultraslow}. The most promising molecular and stoichiometric materials studied to date exhibit homogeneous linewidths that range from 0.5~kHz to 0.5~MHz \cite{ahlefeldt2013optical, ahlefeldt2016ultranarrow, shelby1980frequency, serrano2022ultra}.

\begin{figure}[h]
    \centering
    \includegraphics[width=0.47\textwidth]{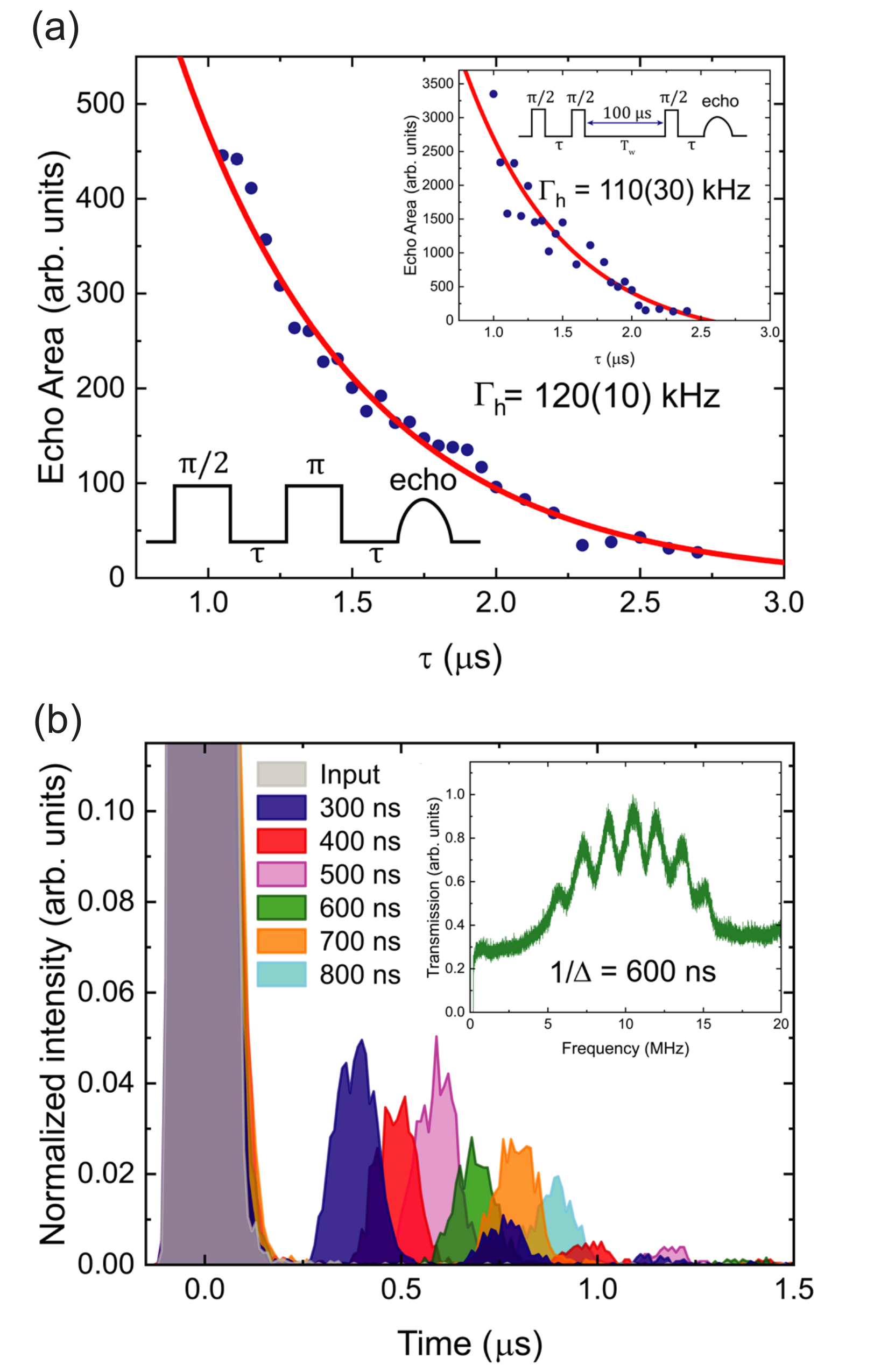}
    \caption{Coherence properties of \ce{^5D_0}$\rightarrow$\ce{^7F_0} transition \ce{NaEu(IO3)4} at 1.7~K and preliminary AFC storage and retrieval. (a) Photon echo studies for a two-pulse photon echo and a three-pulse photon echo for a wait time of $T_w = 100~\mu\rm{s}$ (inset). The decay of the echo amplitudes as a function of the delay $\tau$ are fitted to a single exponential given by the red curve yielding a decay constant $\tau_{decay}$ which yields a coherence time $T_2 = 4\tau_{decay}$. (b) Preliminary demonstrations of coherent storage of bandwidth-match pulses in 8.3~MHz AFCs for different storage times. The amplitudes of the retrieved echo pulses were normalized to their respective input pulses. For storage times of $t_{s}$ = 300~ns, 400~ns, and 500~ns, multiple echo pulses can be seen (indigo, red, and magenta echo pulses, respectively). The inset depicts a typical AFC with a $1/\Delta$ = 600~ns, color matched to the main figure.}
    \label{fig:coherence}
\end{figure}

To directly measure the optical coherence time (and thus the homogeneous linewidth $\Gamma_h=1/\pi T_2$), we performed a photon echo sequence \cite{macfarlane1992laser, thiel2011rare,kaplyanskii2012spectroscopy}. A short (500~ns) pulse creates coherence between the ground and excited state which quickly dephases due to the inhomogeneous spread of atomic energy levels. Another short pulse after a delay time $\tau$ rephases the ensemble, leading to the re-formation of the initial macroscopic coherence and the emission of an ``echo" pulse. This rephasing will occur as long as the emitters remain coherent between the input pulse and the echo, and thus, by measuring the decay of the echo with $\tau$, we can extract the optical coherence time. The echo relies on momentum conservation, requiring operation in a transmission geometry, so this was implemented with the same powder sample as the hole burning. 
The intensity of the detected echo signal is shown as a function of delay time $\tau$ between the centers of the first and second pulses in Figure \ref{fig:coherence}. To quantify the emitter coherence time, we use the expression 

$$ I(\tau) = I_0 \exp\left(\frac{-4\tau}{T_M}\right)^x $$

where the parameter $x$ determines the shape of the decay and characterizes the spectral diffusion \cite{mims1968phase}. $T_M$ is the phase memory time and in the limit $x =1$, the phase memory is the optical coherence time: $T_M = T_2$ \cite{thiel2011rare, ganem1991nonexponential}. We observe the echo intensity to be well-fit by a single exponential decay with a time constant of $\tau_{\rm{decay}}=0.6(1)~\mu\rm{s}$, which corresponds to an optical coherence time $T_2=4\tau_{\rm{decay}}=2.4(4)~\mu\rm{s}$ and an optical homogeneous linewidth $\Gamma_h=1/\pi T_2=0.12(1)~\rm{MHz}$. 

Three-pulse or stimulated photon echo techniques have been used as a sensitive probe for spectral diffusion, dephasing mechanisms that occur on a timescale longer than $T_2$ (but shorter than the lifetime, $T_1=1.5~\rm{ms}$) \cite{yano1992stimulated, kunkel2016dephasing, bottger2006optical, goldner2015rare, levenson2012introduction}. The first two pulses generate a frequency-dependent diffraction grating that stores the accumulated phase of the coherence between the ground and excited state. A third pulse is applied after some intermediate waiting time, $T_w$, which diffracts off of the grating and rephases the coherence. A stimulated echo is emitted at a time $\tau$ after the third pulse. For a fixed $T_w$, the decay of the echo intensity as a function of $\tau$ will yield an effective homogeneous linewidth $\Gamma_{eff,T_w}$, which will broaden for increasing $T_w$ in the presence of spectral diffusion. We implemented a stimulated echo for a fixed waiting time of $T_w = 100~\mu\rm{s}$ (see the inset of Fig. \ref{fig:coherence}a) and observed $\Gamma_{eff, 100~\mu\rm{s}}$ = 110(30)~kHz, within the uncertainty of the homogeneous linewidth at zero wait time. We note that in the absence of spectral diffusion, the decay of the initial echo intensity as a function of $T_w$ should yield an exponential decay with a decay constant equal to the population decay rate, $T_1$, enabling observation of $\Gamma_{eff, T_w}$ out to wait times nearing the lifetime. However, we were unable to obtain sufficiently large echo signals for $T_w>100~\mu\rm{s}$ to determine this, Thus, we cannot conclusively rule out the impact of spectral diffusion on the coherence properties on a timescale longer than 100~$\mu\rm{s}$. 

Finally, we demonstrated preliminary measurements of retrieval of echo pulses using atomic frequency combs (AFCs) as a delay line. The AFC protocol is a quantum memory protocol that has been shown to efficiently and faithfully store single photonic qubits \cite{clausen2011quantum}. We prepare a periodic comb structure in the ensemble absorption profile via spectral hole burning. 
A resonant input signal pulse with bandwidth much larger than the comb tooth separation $\Delta$ is absorbed, generating a collective excitation of the emitters that form the comb and a coherence that rapidly dephases due to the inhomogeneity of the ion’s transition frequency. However, the periodic structure of the comb causes the ensemble to periodically rephase at a time $1/\Delta$ and emit an echo pulse. Thus, the delay is then set by the spacing of the comb teeth, $t_{delay} = 1/\Delta$\cite{afzelius2009multimode, simon2010quantum, shinbrough2023broadband}. 

We prepared AFCs with a bandwidth of 8.3~MHz for storage times of 300~ns to 800~ns. Storage experiments were then performed with bandwidth-matched, weak, coherent input pulses. Retrieval of the emitted echo pulses can be seen in Fig. \ref{fig:coherence}b, where we observed echoes at the expected storage time after the input pulse with a maximum storage efficiency of $\approx2\%$ for 300~ns storage and an efficiency of 0.8\% for a storage time of 800~ns. In the case of $t_{delay}$ = 300~ns, 400~ns, and 500~ns, the emission of multiple echo pulses can be seen after another $t_{delay}$ from the preceding echo. These exploratory measurements and the demonstration of long-lived, deep spectral holes are promising for utilizing NaEu(IO$_3$)$_4$ as an efficient spin-wave quantum memory. 

In conclusion, we have demonstrated narrow optical linewidths in the 2D-layered stoichiometric material, \ce{NaEu(IO_{3})_{4}}. The inhomogeneous linewidth of the $^5$D$_0$$\rightarrow$$^7$F$_0$ transition is comparable to Eu$^{3+}$-doped materials of low doping concentration, and the homogeneous linewidth is similar to other molecular materials recently investigated for QIP applications. 
Our observation of spectral holes with lifetimes in excess of minutes enables the possibility of efficient optical spin polarization for long-lived quantum memory. Furthermore, while the QIP parameters of interest in \ce{NaEu(IO_{3})_{4}} are comparable to other stoichiometric material platforms in the literature \cite{kumar2021optical, kuppusamy2023observation, serrano2022ultra, schlittenhardt2024spectral}, the 2D-layered nature presents a promising avenue for potential integration with nanophotonic devices for a scalable light-matter interface. 

\section*{Methods}

Synthesis of the two batches of \ce{NaEu(IO_{3})_{4}}, experimental diagrams and descriptions, and further details of the SHB and photon echo pulse sequences can be found in the Supplementary Information. Additional details, such as part numbers, are also addressed in further detail in the Supplementary Material.

\subsection{High-resolution optical spectroscopy}

All cryogenic measurements were conducted in a Janis Supertran-VP liquid helium flow cryostat at 1.7~K. PL spectra were collected in a confocal setup under 405~nm illumination. The emitted fluorescence was separated from the excitation light by a 550~nm long-pass dichroic mirror and further filtered by a 495~nm long-pass filter before being coupled into a multimode fiber connected to a visible spectrometer (Ocean Insights QE Pro). The PL  lifetime was measured using resonant excitation from a TOPTICA DL SHG-PRO frequency-doubled diode laser.

Optical pulses were generated using an acousto-optic modulator (Isomet, M1206-P110L-1). The excitation light was separated from the fluorescent emission with a 593~nm long-pass filter, and the fluorescent emission was collected into a multimode fiber connected to a single photon avalanche diode (SPCM-AQRH-14-FC, Excelitas Technologies). PLE was conducted using the frequency-doubled diode laser. The frequency of the $\approx$ 580~nm excitation light was modulated by scanning the current of the 1160~nm master diode and was monitored via a pick-off of the 1160~nm to a wavemeter (Bristol 761). \ce{NaEu(IO_{3})_{4}} crystals were mounted in the cryostat with copper tape to custom copper sample plates. Fluorescence was collected on the SPAD, and the inhomogeneous linewidth was then stitched together over several overlapping scans and fitted to a Lorentzian. 

\subsection{Spectral hole burning}

Spectral hole burning was conducted using a transmission geometry with an $\approx$ 100~kHz-linewidth laser stabilized to a low finesse reference cavity using a side-of-fringe lock (Sirah Matisse CR with a mix-train module). A double-passed AOM was used to generate the pump and probe beam pulses. The transmitted speckle pattern was then sent to a short-pass filter to eliminate the fluorescent emission and focused onto an avalanche photodiode (Thorlabs APD120A). 

\subsection{Coherence properties and preliminary light storage}

Two-pulse photon echo measurements were conducted using 500~ns $\pi/2$ and $\pi$-pulses. To avoid hole burning, the laser was stepped in frequency between each sequence. The transmitted speckle pattern was focused through an AOM for temporal gating and collected into a multimode fiber connected to a SPAD. Fluorescent emission was separated from the echo signal with a short-pass and a 580~nm band-pass filter. AFCs were created at the center of the inhomogeneous line with a sequence of $10^6$ 120~ns pulses at various repetition rates. Weak 120~ns pulses were then sent and retrieved. 

\section{Acknowledgments}

A.T. was supported by the NSF through the University of Illinois at Urbana-Champaign Materials Research Science and Engineering Center DMR-2309037 via the Research Experiences for Undergraduates (REU) program. S.T. was supported by an Open Quantum Initiative Undergraduate Fellowship from the Chicago Quantum Exchange. This material is based upon work supported by the U.S. Department of Energy, Office of Science, National Quantum Information Science Research Centers.


\bibliographystyle{ieeetr}

\pagebreak
\widetext
\begin{center}
\textbf{\large Supplemental Information for Narrow optical linewidths in stoichiometric layered rare-earth crystals}
\end{center}
\setcounter{equation}{0}
\setcounter{figure}{0}
\setcounter{page}{1}
\makeatletter
\renewcommand{\theequation}{S\arabic{equation}} 
\renewcommand{\thefigure}{S\arabic{figure}}

\section{Materials}
\subsection*{Preparation of \ce{NaEu(IO3)4} single crystals and powder}

Two samples of \ce{NaEu(IO3)4} (ST009 and AT006-1) were prepared by solvothermal reaction of Eu(NO$_{3}$)$_{3}$$\cdot$6H$_{2}$O and \ce{NaIO3} in a Parr 4749 acid digestion vessel containing  1.26 M  HNO$_{3}$. The resulting crystals were washed with deionized water after decanting the solvent (Figure S1). Representative portions of each sample were ground for powder x-ray diffraction in a Brucker D8 diffractometer, with phases assigned by Rietveld refinement (Figure S2 and S3)

ST009 was prepared by addition of 0.4566~g of Eu(NO$_{3}$)$_{3}$$\cdot$6H$_{2}$O (Thermo Scientific, 99.99\%) to 10~mL of deionized water, followed by addition of 1.9797~g of \ce{NaIO3} (Thermo Scientific, 99.99\%) and 0.857~mL of HNO$_{3}$ (Fisher Scientific, 70\%). The reactant mix was transferred to the acid digestion vessel. The vessel was heated to 200~\degree C at a rate of 20~\degree C per minute, remained at 200~\degree C for 72 hours, and then cooled at a rate of 6~\degree C per hour.

AT006-1 was prepared by addition of 0.4445~g of  Eu(NO$_{3}$)$_{3}$$\cdot$6H$_{2}$O (Thermo Scientific, 99.99\%) with 1.9792 g \ce{NaIO3} (Thermo Scientific, 99.99\%) in 10~mL of 1.26~M \ce{HNO_{3}}. The reactant mix was transferred to the acid digestion vessel. The vessel was heated at 200~\degree C for 2 hours before cooling to 30~\degree C at 4~\degree C per hour. 
\begin{figure}
    \centering
    \begin{subfigure}{0.5\linewidth}
        \centering
        \includegraphics[width=\linewidth]{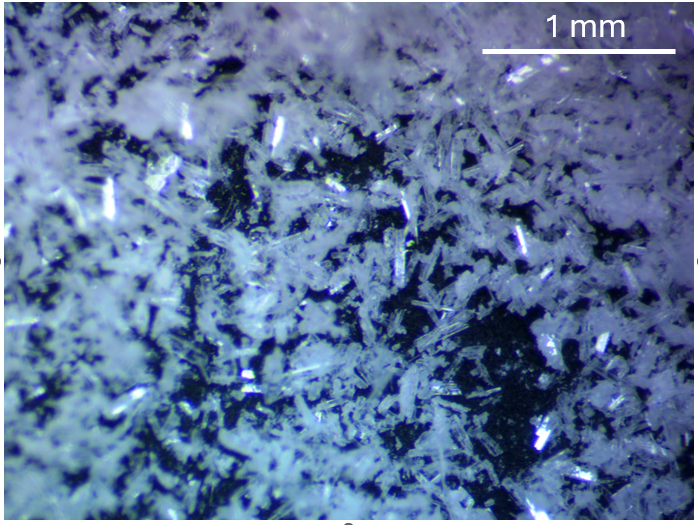}
    \end{subfigure}%
    \begin{subfigure}{0.5\linewidth}
        \centering
        \includegraphics[width=\linewidth]{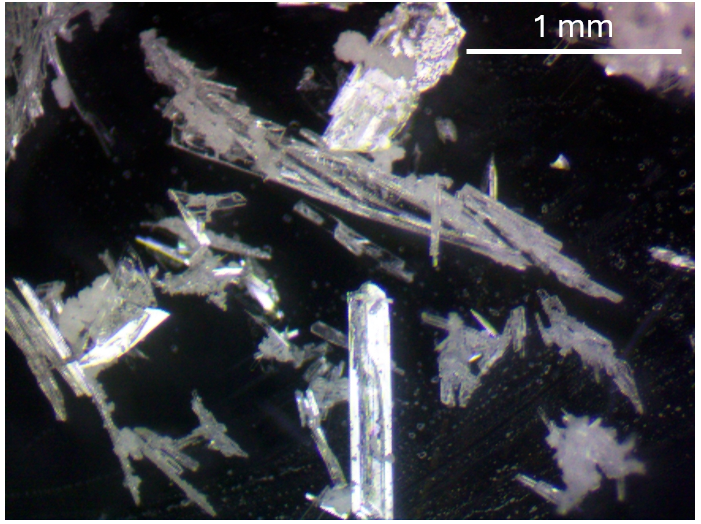}
    \end{subfigure}
    \caption{Crystals from ST009 (left) and AT006-1 (right).}
    \label{fig:AuSbTe_struture}
\end{figure}
\newpage

\begin{figure}
    \centering
    \includegraphics[width=0.8\linewidth]{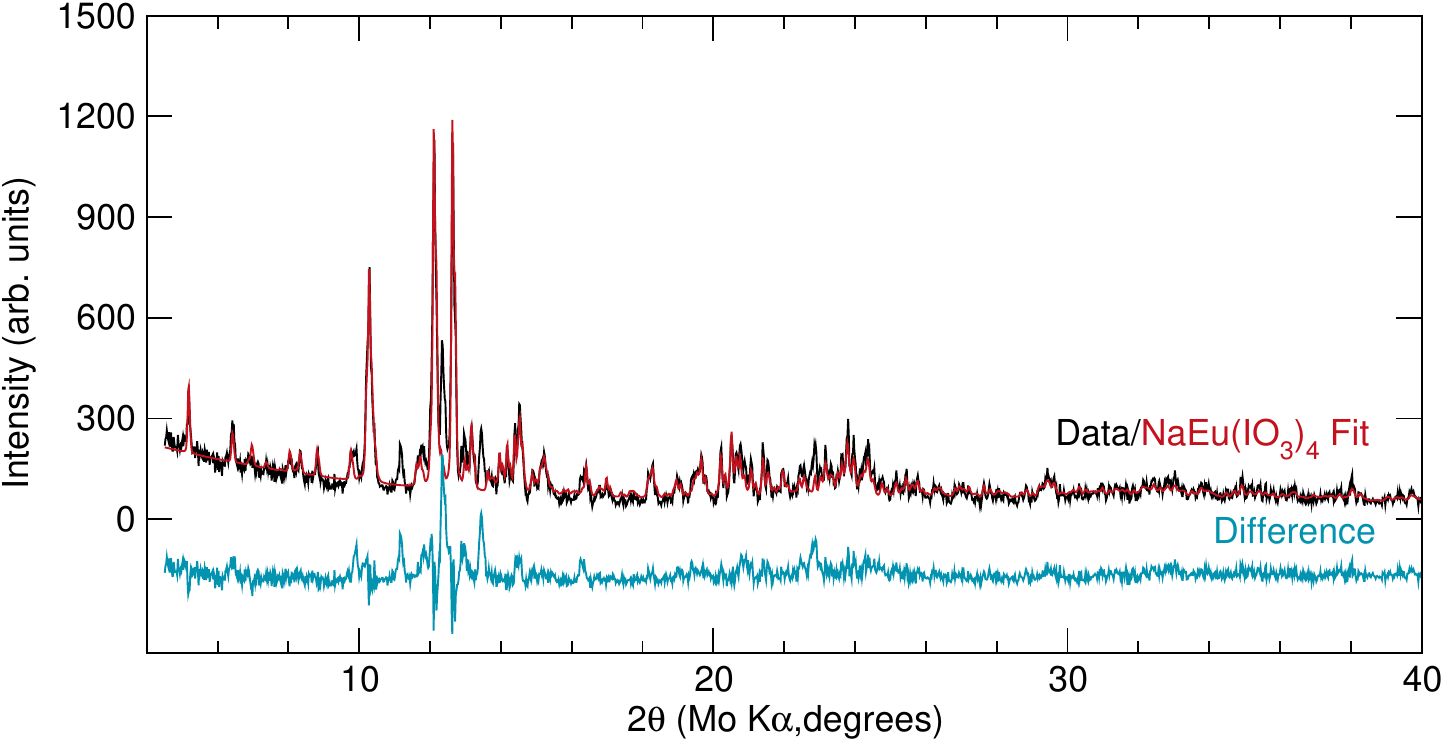}
    \caption{Rietveld refinement of room temperature powder pattern of AT006-1 (black) to predicted pattern of \ce{NaEu(IO3)4} (red). The minor impurity was not positively identified. }
    \label{a}
\end{figure}
\begin{figure}
    \centering
    \includegraphics[width=0.8\linewidth]{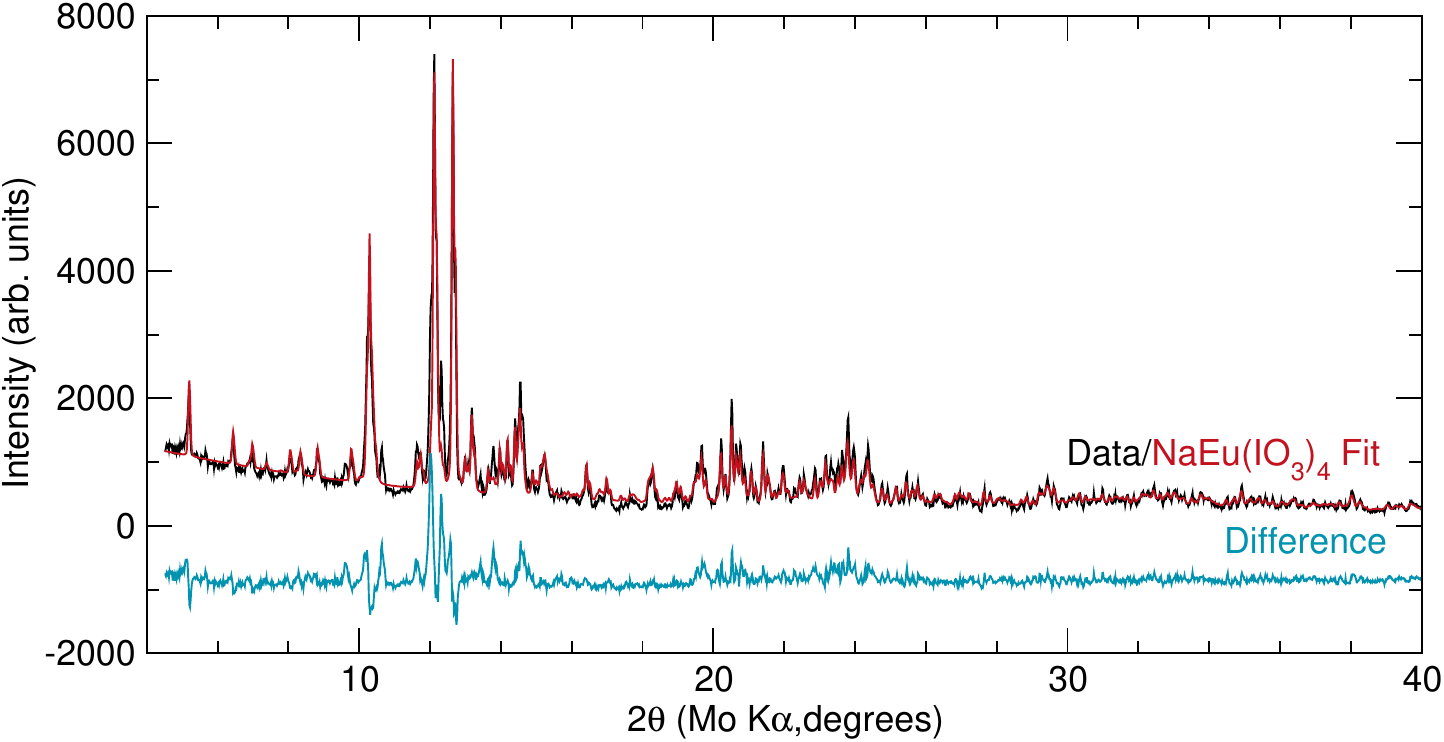}
    \caption{Rietveld refinement of room temperature powder pattern of ST009 (black) to predicted pattern of \ce{NaEu(IO3)4} (red). The minor impurity was positively identified as \ce{Eu(IO3)3.}}
    \label{a}
\end{figure}
Crystals from ST009 were used for photoluminescence-excitation spectroscopy with the detection of fluorescent emission. Samples from AT006-1 were ground into a powder and used for photoluminescence, spectral-hole burning in transmission, and photon echo studies.

\section{Room Temperature Photoluminescence}

\begin{figure}
    \centering
    \includegraphics[width=0.8\linewidth]{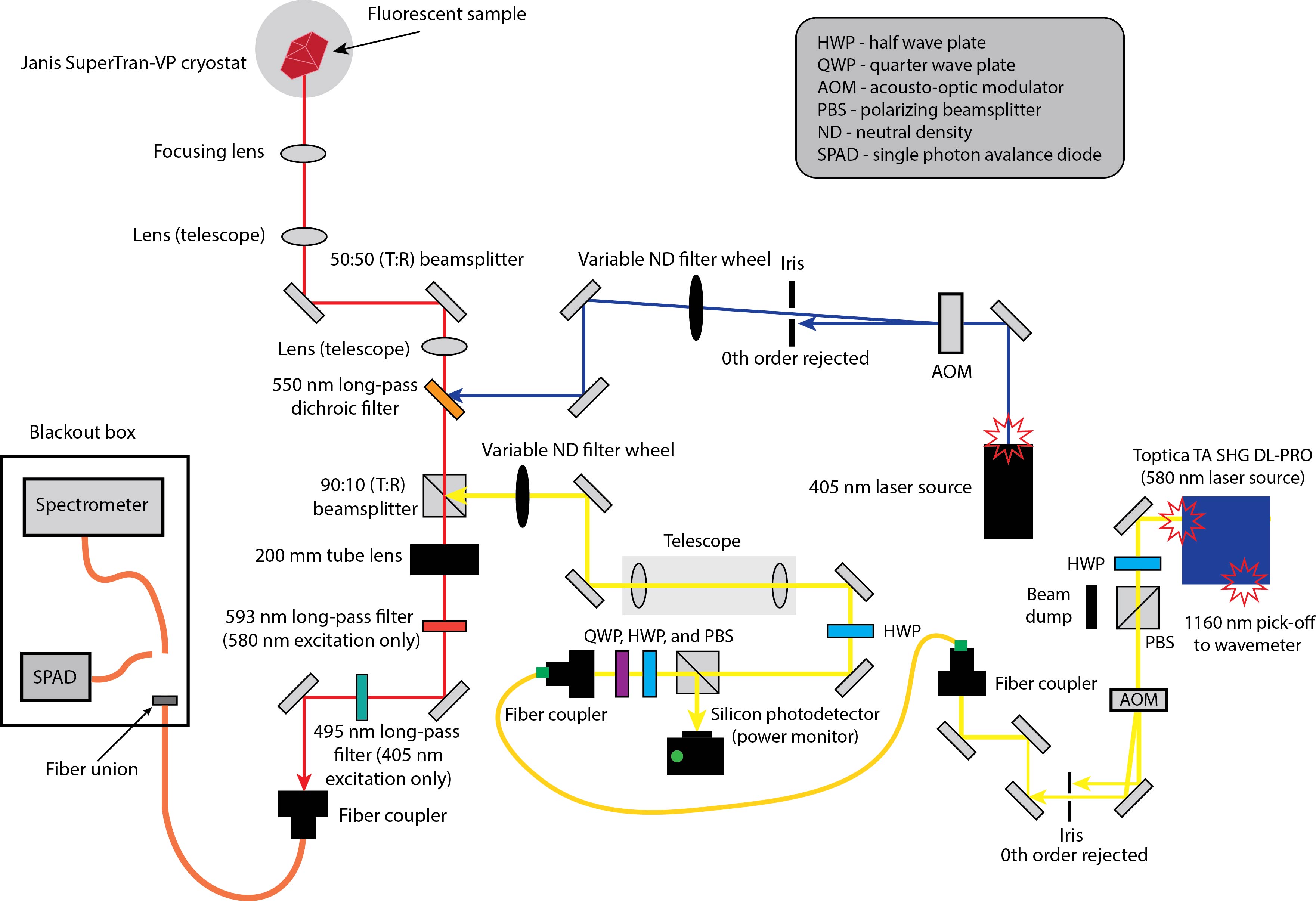}
    \caption{Schematic for PL and PL lifetime measurements at room temperature and 1.7 K}
    \label{fig:PL&lifetimesetup}
\end{figure}

\begin{figure}
    \centering
    \includegraphics[width=\linewidth]{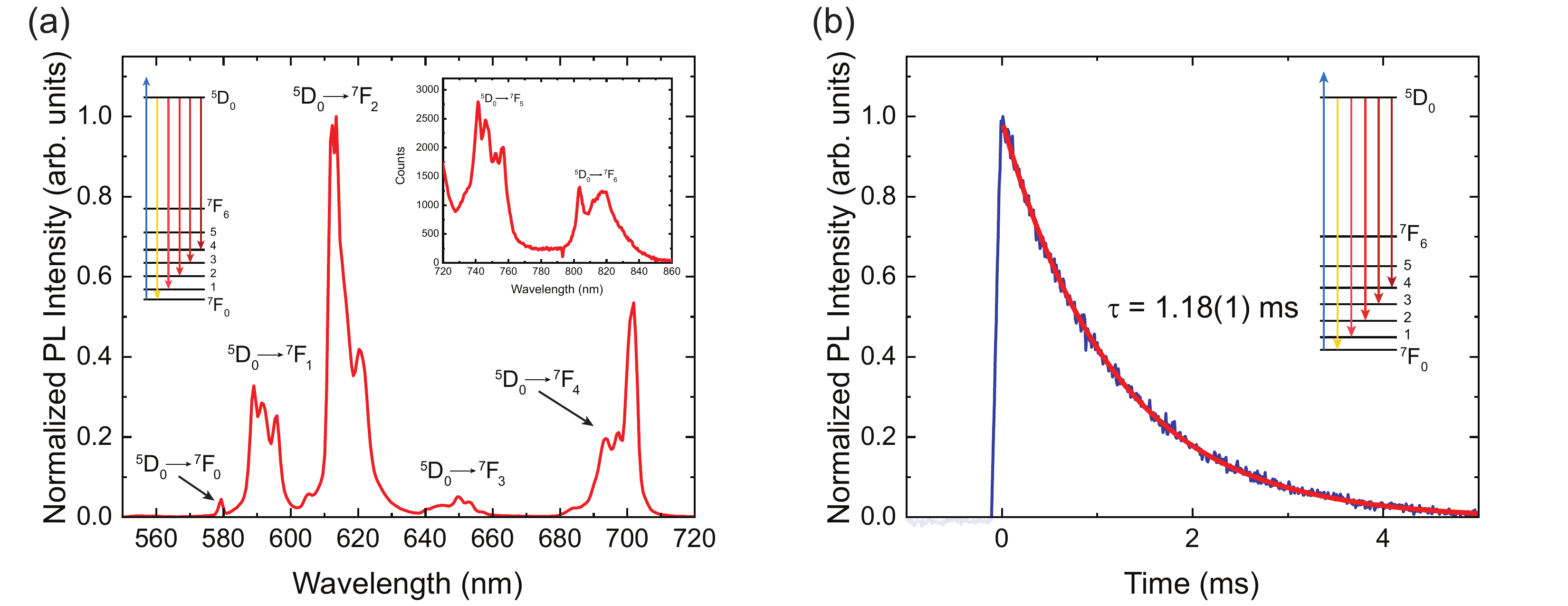}
    \caption{Room temperature data photoluminescence of NaEu(IO3)4. (a) Photoluminescence spectrum collected with 405 nm light. The J = 5 and J = 6 transitions can be seen in the inset. The intensity is not normalized for the inset due to the saturation of the spectrometer. (b) The photoluminescence lifetime using 100-microsecond pulses of 405~nm excitation light. }
    \label{fig:roomtempPL&lifetime}
\end{figure}

Room temperature photoluminescence (PL) and photoluminescence lifetime measurements were taken using a confocal setup (see Fig.~\ref{fig:roomtempPL&lifetime}) on the second batch of NaEu(IO$_3$)$_4$. PL measurements were conducted with 405~nm excitation light. Pulses for the PL lifetime were generated with an acousto-optic modulator (AOM) from Gooch \& Housego (3100-125). The excitation light was sent to the sample via the rejection of a 550~nm long-pass dichroic mirror (Thorlabs DMLP550). The fluorescent emission was transmitted through the dichroic and further filtered with a 495~nm long-pass filter (Thorlabs  FGL495M)  before being coupled into a multi-mode fiber connected to the appropriate detector. All PL data was collected with a fiber-coupled Ocean Insights QE Pro visible spectrometer, and PL lifetime data was collected with a single photon counting module (SPCM-AQRH-14-FC, Excelitas Technologies). 

The PL and PL lifetime at room temperature of NaEu(IO$_3$)$_4$ have been reported in the literature \cite{oh2017photoconversion}. We additionally report the observation of the $^5$D$_0$~$\rightarrow$~$^7$F$_0$ transitions for $J$ = 0, 3, 5, and 6. The spectrum is shown in Fig.~\ref{fig:roomtempPL&lifetime}(a). The $J$ = 5 and 6 features were observed with optical excitation powers that saturated the spectrometer and were not normalized to the rest of the spectrum (see inset of Fig.~\ref{fig:roomtempPL&lifetime}(a)). We conducted room temperature PL lifetime measurements with 100~$\mu\rm{s}$ pulses of 405~nm excitation light. The signal was collected on a single photon counting module and integrated for 120 seconds using 600, 10~$\mu\rm{s}$ time bins. The resulting fluorescent decay was well-fitted to a single exponential which was fitted from 30~$\mu\rm{s}$ after the pulse turns off ($t$ = 0) through the end of the collection window (5 ms). The room temperature lifetime was observed to be $T_{1,opt}$ = 1.18(1)~ms. This can be seen in Fig.~\ref{fig:roomtempPL&lifetime}(b). This is longer than the value reported in Oh. et al. \cite{oh2017photoconversion}. 

\section{Cryogenic Temperature Photoluminescence Spectroscopy \& Photoluminescence Lifetime}

Experiments at superfluid helium temperatures were conducted in a Janis SuperTran-VP liquid helium bath cryostat. The temperature was detected using a silicon diode sensor (DT-670-CU-HT-1.4L, LakeShore) mounted on the sample rod and monitored with a LakeShore 240-2P module. Photoluminescence at 1.7~K was obtained using the same setup as the room temperature measurement in Fig.~\ref{fig:PL&lifetimesetup}. The collected PL spectrum was obtained by integrating the collected fluorescent emission for 30 seconds. From the spectrum in the main text, the branching ratio of the $^5$D$_0$~$\rightarrow$~$^7$F$_0$ transition was obtained by integration of the spectrum. The oscillator strength was then calculated using the branching ratio, assuming an index of refraction $n$ = 1.5, the optical lifetime at 1.7~K, and the center frequency of the $^5$D$_0$~$\rightarrow$~$^7$F$_0$ transition. 

The lifetime at 1.7~K was obtained using 50~$\mu\rm{s}$ pulses on the resonance peak with a frequency-doubled diode laser (Toptica TA DL pro SHG) set to 516,960.21(6)~GHz. Pulses were generated with an AOM (Isomet, M1206-P110L-1). The laser was actively stabilized using a Pound-Drever-Hall lock for all measurements in which it was used. The pulse sequence consisted of a 948~$\mu\rm{s}$ delay followed by a 50~$\mu\rm{s}$ pulse. After the pulse, there is a 2~$\mu\rm{s}$ delay to account for the AOM turning on and off. This is then followed by 5~ms of time to observe the fluorescent emission; the beginning of this observation time is taken to be $t$ = 0. The emitted fluorescence was collected with 600, 10~$\mu\rm{s}$ time bins. The signal was then integrated for 20~mins, and a background was subtracted. The signal was then fit to a single decaying exponential as detailed in the main text. The oscillator strength was measured by integrating the 1.7~K PL spectrum to obtain the branching ratio. The spectrum was integrated from 577~nm to 850~nm to capture the entirety of the $^5$D$_0$~$\rightarrow$~$^7$F$_J$ emission.

\section{Photoluminescence Excitation Spectroscopy at 1.7~K}

\begin{figure}
    \centering
    \includegraphics[width=0.75\linewidth]{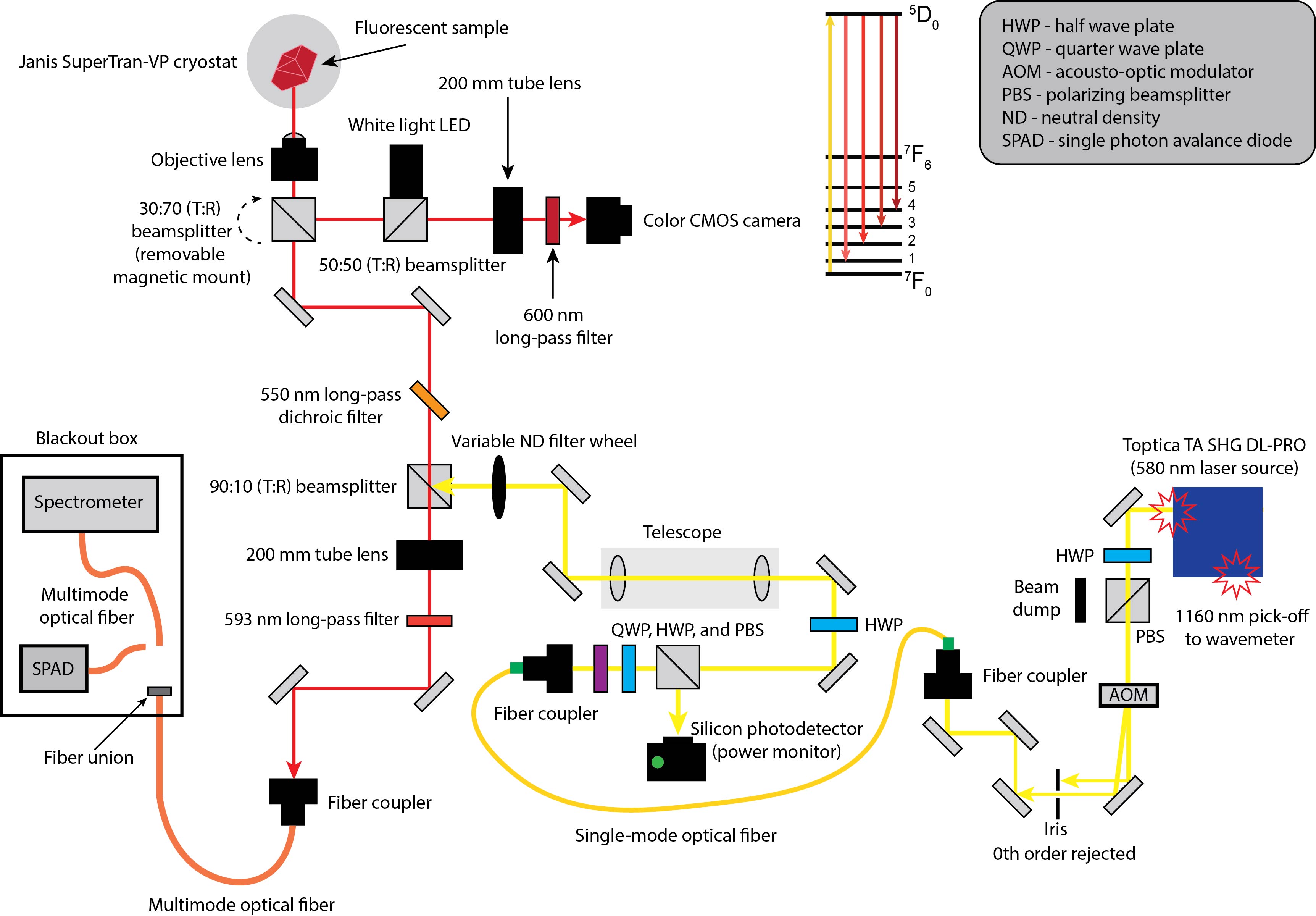}
    \caption{Schematic diagram for photoluminescence excitation spectroscopy of single crystals from the first batch of NaEu(IO3)4. Inset is the relevant energy level diagram with 580 nm excitation and collection of the red-detuned fluorescent emission. }
    \label{fig:PLEsetup}
\end{figure}
The inhomogeneous linewidth of the $^5$D$_0$~$\rightarrow$~$^7$F$_0$ transition was measured at 1.7~K using photoluminescence excitation spectroscopy (PLE) in a confocal geometry shown in Fig.~\ref{fig:PLEsetup} for single crystals and Fig.~\ref{fig:PL&lifetimesetup} for the powder samples.  In the setup shown in Fig.~\ref{fig:PLEsetup} for measuring the single crystals, a 30:70 (T:R) beamsplitter sent some of the fluorescent emission to a color CMOS camera (AMscope MD310C-BS) to image the fluorescent emission of the single crystals and ensure we were aligned. Samples were mounted on copper tape and connected to a stack of two single-axis, low-temperature nanopositioner stages (attocube ANPx51 and ANPz51) with a custom sample holder so the crystals could be aligned with our imaging system. The beam was focused on single crystals using a x2 magnification finite conjugate objective lens with an effective focal length of 60.19~mm and a working distance of 92~mm (Edmund Optics 56-989, manufactured by Mitutoyo). 

The frequency-doubled diode laser was swept across the 580~nm transition peak via modulation of the master diode current. The frequency of the light was monitored using a pick-off of the 1160~nm fundamental light with a Bristol 761 wavemeter. The emitted fluorescence was isolated from the excitation light by transmission through a 593~nm long-pass filter (Semrock FF01-593/LP-25) before being coupled into a multi-mode optical fiber connected to a SPAD. The collected spectra were stitched together and fitted with a Lorentzian function. The uncertainty on the FWHM was calculated using error propagation arising from the uncertainty on the wavemeter (60~MHz) and the scan period for each scan. The Lorentzian fit uncertainty was then added in quadrature. The center frequency and its uncertainty reported are the mean and standard deviation taken over the fits of eleven different PLE spectra. The same experimental setup shown in Fig.~\ref{fig:PL&lifetimesetup} was used to collect the PLE on the powder samples from the second batch, which were used for photon echo and spectral hole burning studies. 

\section{Spectral Hole Burning Studies at 1.7~K}

\begin{figure}
    \centering
    \includegraphics[width=1.0\linewidth]{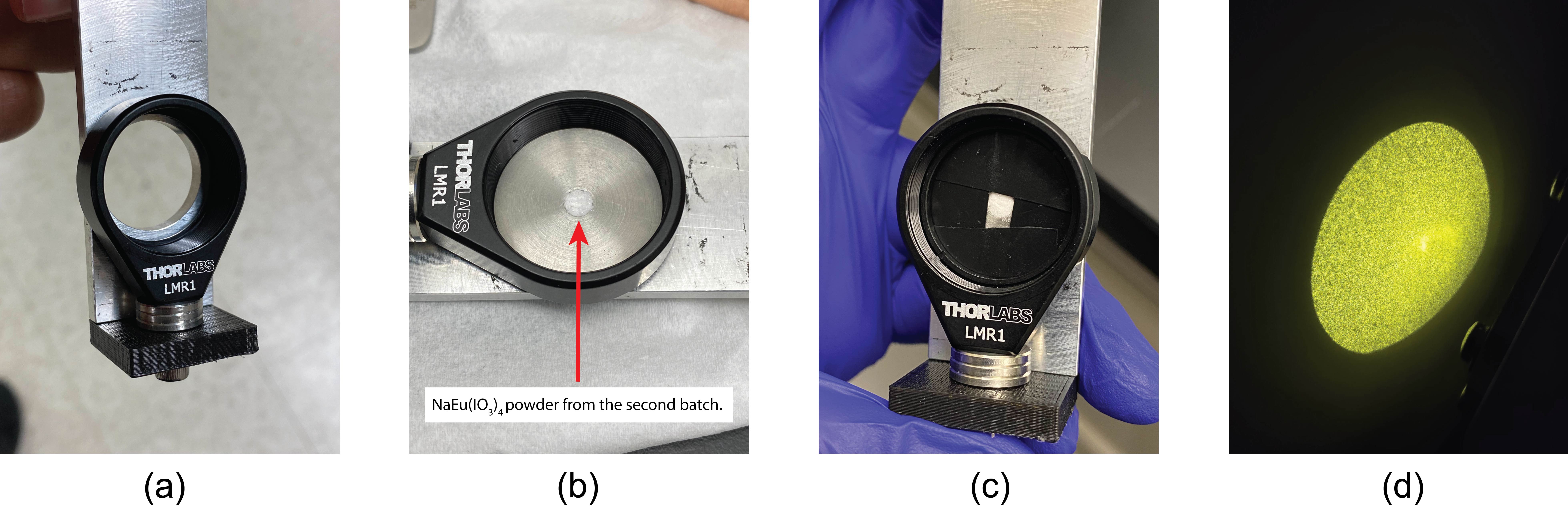}
    \caption{Transmission sample holder and speckle pattern. (a) The sample holder mount. (b) The NaEu(IO3)4 powder packed into the sample holder before the top sapphire wafer was mounted and secured with a retaining ring. (c) The final configuration of the sample mount with the powder packed between the transparent sapphire wafers and then covered in black electrical tape to act as a mask. (d) A typical transmission speckle pattern we observed while aligning the incident beam through the powder sample.}
    \label{fig:transmissionholder}
\end{figure}

\begin{figure}
    \centering
    \includegraphics[width=0.8\linewidth]{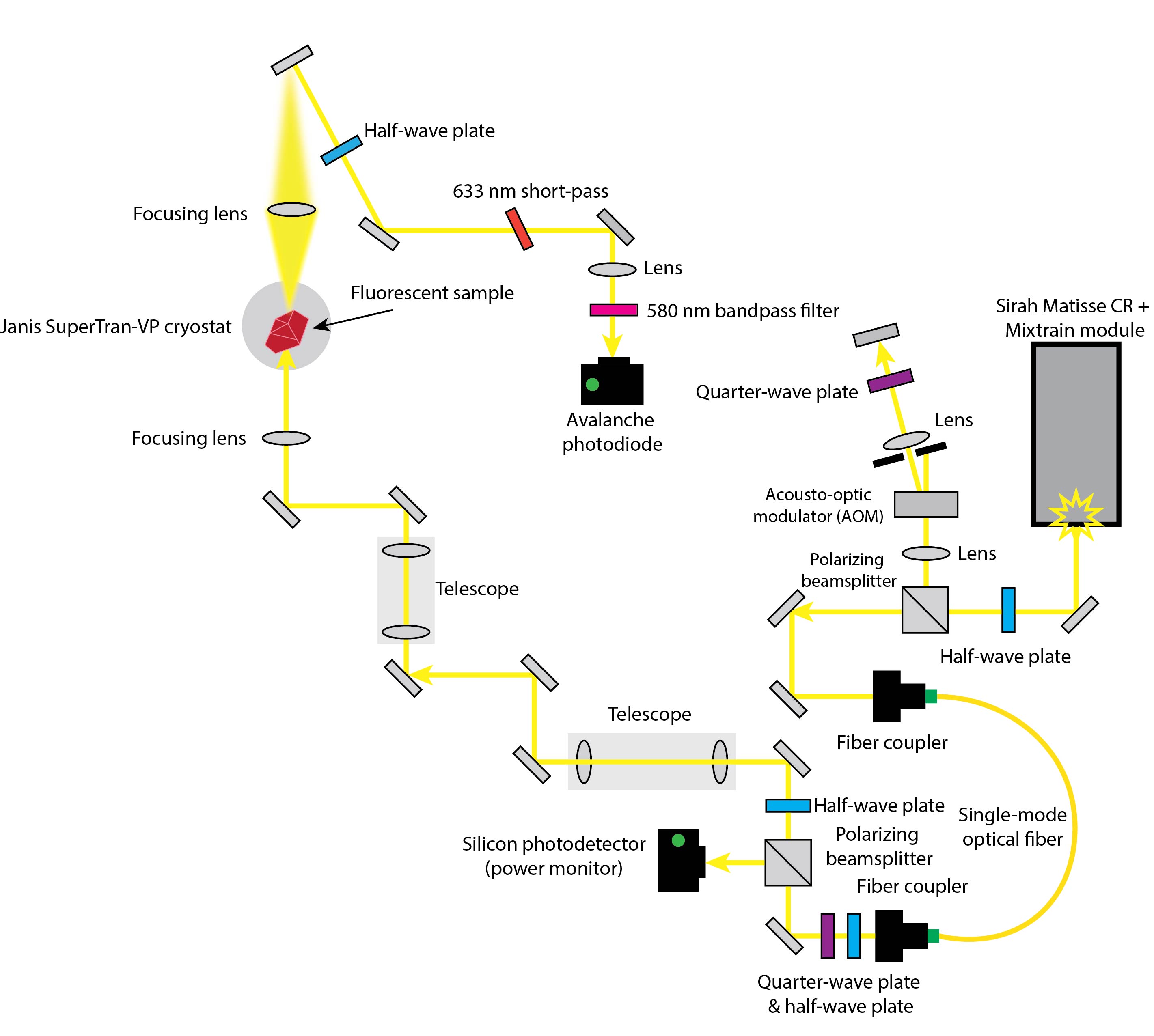}
    \caption{Transmission geometry experimental diagram for the spectral hole burning measurements at 1.7 K.}
    \label{fig:spectralholeburningsetup}
\end{figure}

Spectral hole burning was conducted using a transmission geometry with a $\approx$ 100~kHz laser (Sirah Matisse CR with a mix-train module). The modified beam path can be seen in Fig.~\ref{fig:spectralholeburningsetup}. Samples were mounted in a 1-inch lens mount (Thorlabs LMR1) (see Fig.~\ref{fig:transmissionholder}(a)). The crushed powder from the second batch was placed in a small volume of a disk with an aperture of 4 mm and a thickness of 0.75~mm (see Fig.~\ref{fig:transmissionholder}(b)). This was then sandwiched between two circular, transparent sapphire wafers. Each sapphire wafer has a thickness of 600~$\mu\rm{m}$. This resulted in an optical depth of $\approx$ 2 at the center of the inhomogeneous profile and $\approx$ 1.5 at 2~GHz off-resonance. Black electrical tape was applied to both sides to act as a mask for the scattered light (see Fig.~\ref{fig:transmissionholder}(c)). A double-passed AOM was used to generate the pump and probe beam pulses (Isomet M1206-P110L-1). The transmitted speckle pattern was then sent to a short-pass filter to eliminate the fluorescent emission. The rejected resonant light was focused on an avalanche photodiode (Thorlabs APD120A). A typical speckle pattern while aligning through the sample holder can be seen in Fig.~\ref{fig:transmissionholder}(d). The pulse sequence consisted of an 100~ms burn pulse followed by a probe that was swept 6 MHz across the spectral hole in 102~$\mu\rm{s}$. The resulting spectral hole was fit to a Lorentzian, and the approximate homogeneous linewidth was given by $2\Gamma_h= \Gamma_{hole}$. This represents an upper bound on the achievable homogeneous linewidth. The uncertainty on the FWHM is simply the fit uncertainty.  

The maximum transmission was measured by burning through the sample with a continuous wave beam and detecting it with the avalanche photodiode. A pick-off of the incident light was used to monitor the input power. The maximum depth feature burned was measured 2~GHz off-resonance with a 1-second-long burn pulse. The depth was extracted from the data normalized to the maximum transmission. 

\section{Photon echo measurements \& preliminary AFC retrieval at 1.7 K}

\begin{figure}
    \centering
    \includegraphics[width=0.8\linewidth]{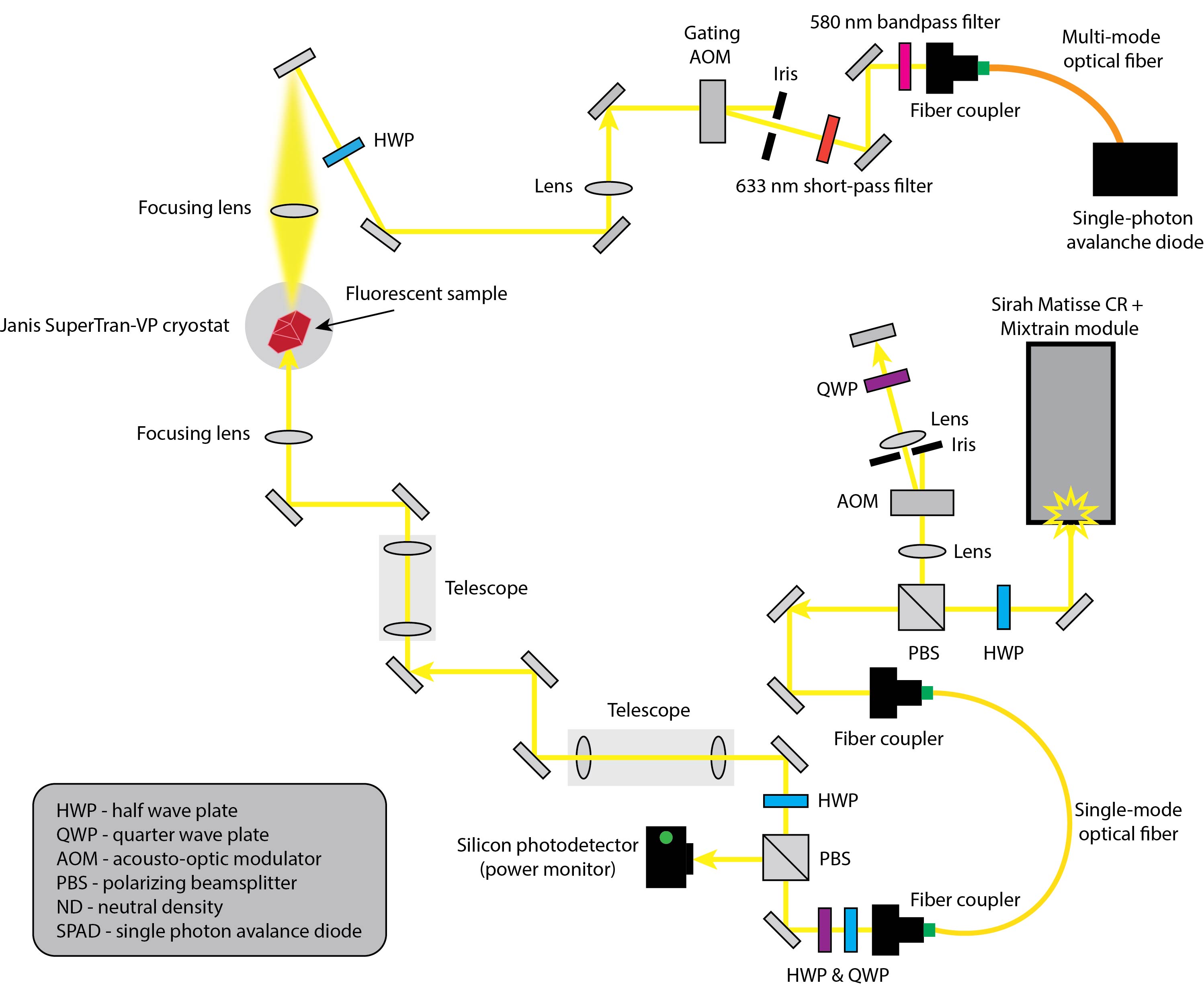}
    \caption{Modified transmission experimental schematic for photon echo measurements and the preliminary atomic frequency comb echo retrieval measurements.  }
    \label{fig:photonechosetup}
\end{figure}

Photon echo measurements were conducted using a modified version of the transmission setup used for the spectral hole burning measurements shown in Fig.~\ref{fig:photonechosetup}. They differ in the detection scheme, which uses a SPAD. A 633~nm short-pass filter (Semrock SP01-633RU-2) was used to reject the fluorescent emission, and a gating AOM (Isomet M1206-P110L-1) to block the bright excitation pulses from getting to the SPAD (see Fig.~\ref{fig:photonechosetup}). A 580~nm$\pm$~10~ bandpass filter (Thorlabs FBH580-10) was added to reject any fluorescence. The collected transmission was focused down to a smaller spot with a telescope and then passed through a gating AOM for temporally filtering out the excitation pulses. Fluorescent emission was separated from the echo signal via a short-pass filter (Semrock SP-01-633RU-2) and the same 580~$\pm$~10~nm bandpass filter. The transmitted light was coupled into a multi-mode fiber connected to a SPAD. 

For the two-pulse photon echo procedure, both the $\pi/2$ and $\pi$-pulses were 500~ns in duration, which was chosen solely because this resulted in the highest signal-to-noise ratio. The three-pulse photon echo measurement with a waiting time of $T_w$ = 100~$\mu\rm{s}$ was taken using pulse durations of 500~ns. To avoid spectral hole burning during the photon echo measurements, the frequency of the laser was stepped for each delay $\tau$ in both two and three-pulse photon echo experiments.

\begin{figure}
    \centering
    \includegraphics[width=0.8\linewidth]{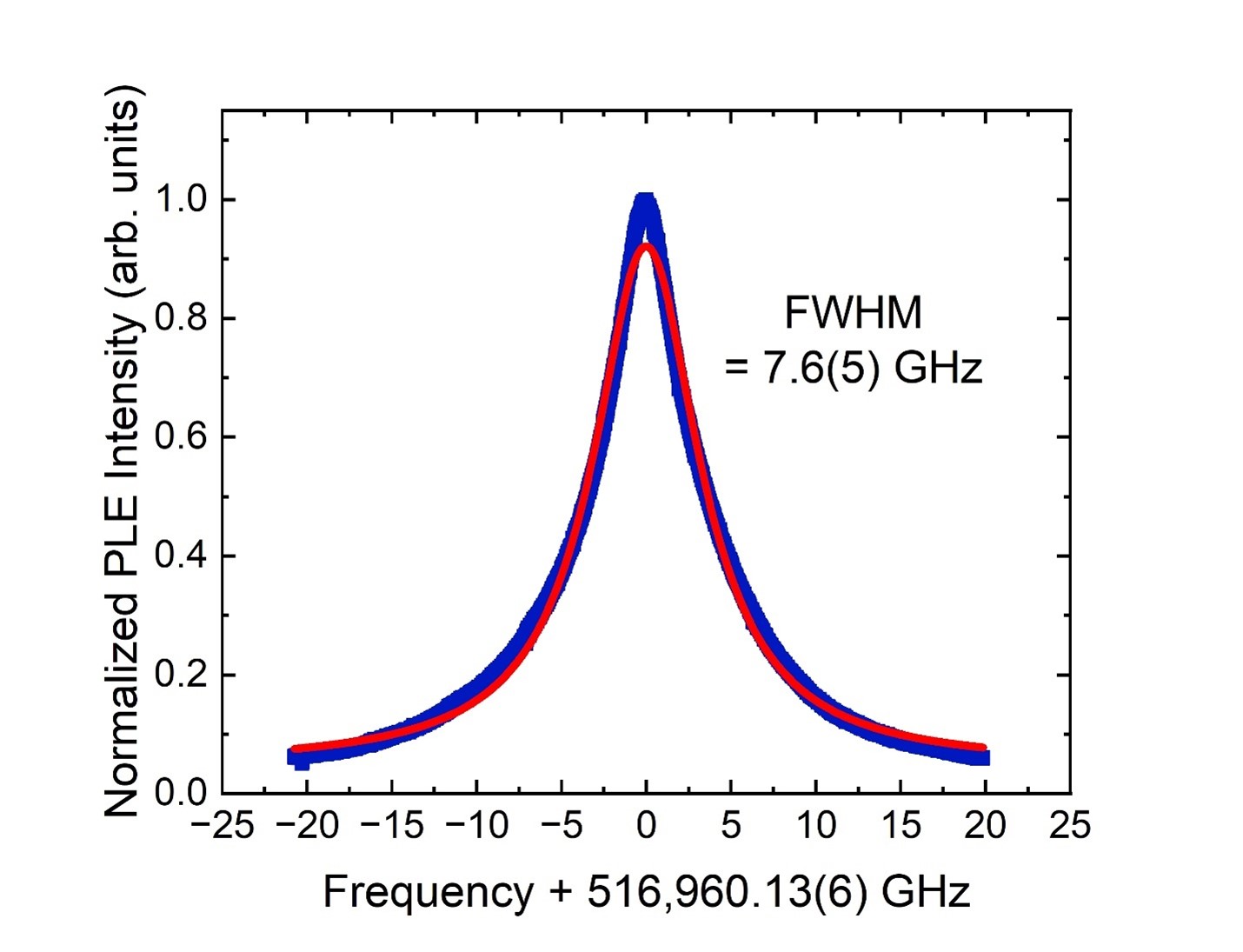}
    \caption{The PLE spectrum of the powder from the second batch measured at 1.7 K using a hybrid of the confocal geometry with the focusing optics of the transmission setup instead of the microscope objective lens. }
    \label{fig:powderPLE}
\end{figure}

To justify stepping in frequency for photon echo measurements, the photoluminescence excitation spectrum was observed for the powder to ensure the inhomogeneous linewidth was broad enough for the optical depth to be approximately the same at the center of the inhomogeneous profile for each frequency step. The inhomogeneous profile of the powder prepared from the second batch of NaEu(IO$_3$)$_4$ is shown in Fig.~\ref{fig:powderPLE}, which yielded an inhomogeneous linewidth of 7.6(5)~GHz FWHM. The spectrum was stitched and fitted to a Lorentzian where the uncertainty reported is the same procedure as previously stated. 

During the collection of the echo, leakage of the $\pi$-pulse (third $\pi/2$-pulse for the three-pulse echo) through the AOM was visible in the detection window of the SPAD to ensure the proper timing of the observed echo pulses. The $\pi$-pulse (third $\pi/2$-pulse for the three-pulse echo) and free-induction decay were filtered from the analysis of the echo area using a window function. The echoes were fit to a Gaussian to acquire the integrated area of the echo for each $\tau$. Decay of the echo area as a function of $\tau$ was fit to a single exponential, and the uncertainty we report for the decay constant $\tau_{decay}$ is given by the fit uncertainty and propagated forward for the coherence time and homogeneous linewidth. 

Atomic frequency combs (AFCs) were prepared using a modified version of the transmission setup in Fig.~\ref{fig:photonechosetup}, which implemented an additional beam shutter due to the bright excitation pulses required to burn the comb structure. AFCs with a bandwidth of 8.3~MHz were prepared at the center of the inhomogeneous profile for a delay time $t_{delay}$ by turning the light on for 120~ns and then off for $t_{delay}$. This was done for $t_{delay}$ ranging from 300~ns to 800~ns. This was followed by probing with a weak, coherent pulse of matching 8.3~MHz bandwidth. The retrieved echo pulses were normalized with respect to the input pulse resulting in a maximum retrieval efficiency ranging from $\approx$ 2\% for 300~ns delay to $0.8\%$ efficiency for 800~ns delay. The retrieved echo pulses were normalized with respect to the input pulses. Since the powdered sample scatters much of the incident light, the input pulse was measured on the SPAD by burning a 20 MHz wide spectral pit. The residual absorption in the pit was determined separately with the APD, and the input pulse was scaled accordingly. The AFC efficiency was measured to range from $2\%$ for 300~ns delay to $0.8\%$ for 800~ns delay. We compare the experimentally measured efficiency of $1.1\%$ for 600~ns delay with the theoretical efficiency. The theoretical efficiency, $\eta_{AFC}$ for a forward propagating echo is\,\cite{afzelius2009multimode}:
\begin{align}
\eta_{AFC}=\bigg(\frac{d}{F}\bigg)^2e^{-\frac{d}{F}} e^{-7/F^2}
\end{align}
where $d$ is the optical depth, and $F$ is the AFC finesse, i.e., the ratio of the teeth spacing to the width of the tooth. The optical depth of the sample and the finesse of the comb were measured to be 2 and 2.32, respectively. The efficiency is then theoretically predicted to be $8.5\%$. The discrepancy between the theoretical and experimental efficiency is due to the poor optical depth of the AFC teeth and residual absorption in the teeth spacing. The efficiency can be improved by using lower power for the AFC preparation pulses and using optimized pulse sequences.

\end{document}